\documentclass[aps,prl,preprint,showpacs,groupedaddress]{revtex4}
\usepackage {amssymb}
\usepackage {amsmath}
\usepackage {graphicx}
\usepackage {longtable}

\begin{document}
\title{The metal-insulator transition in semiconductors}

\author{Fedor V.Prigara}
\affiliation{Institute of Physics and Technology,
Russian Academy of Sciences,\\
21 Universitetskaya, Yaroslavl 150007, Russia}
\email{fvprigara@rambler.ru}

\date{\today}

\begin{abstract}

The temperature dependence of the number density of elementary
excitations in a semiconductor with account for the temperature
dependence of the band gap is obtained. A local lattice distortion
within a crystalline domain is discussed.

\end{abstract}


\maketitle

Recently, it was shown [1] that there is a general relation between the band
gap $E_{g} $ at zero temperature $T = 0K$ and the metal-insulator transition
temperature $T_{MI} $ of the form

\begin{equation}
\label{eq1}
E_{g} = \alpha k_{B} T_{MI} ,
\end{equation}

\noindent
where $k_{B} $ is the Boltzmann constant and $\alpha \approx 18$.

The temperature dependence of the band gap is given by the
equation

\begin{equation}
\label{eq2}
E_{g} \left( {T} \right) = E_{g} - \beta k_{B} T,
\end{equation}

\noindent
where the value of the constant $\beta $ is normally close to 6 ($\beta = 6$
in the case of ZnS and $\beta = 5$ in the case of Si, Ge, and GaAs).

The relation (\ref{eq2}) modifies the temperature dependence of
the number density \textit{n} of elementary excitations [1] in a
semiconductor as follows

\begin{equation}
\label{eq3}
n = n_{0} \left( {T_{MI} /T} \right)^{\beta + 1}exp\left( { - E_{g} /k_{B}
T} \right),
\end{equation}

\noindent
where $n_{0} \approx 1.1 \times 10^{22}cm^{ - 3}$ is a constant. At the
transition temperature $T = T_{MI} $, the equations (\ref{eq3}) and (\ref{eq1}) give the
critical number density of elementary excitations,

\begin{equation}
\label{eq4}
n_{c} = n_{0} exp\left( { - \alpha}  \right) = d_{c}^{ - 3} ,
\end{equation}

\noindent
where $d_{c} \approx 180nm$ is the size of the region around a point defect
in a semiconductor within which there is a local lattice distortion of a
ferroelastic type [2] caused by the charge redistribution. The amplitude
$\delta = \Delta a/a$ (\textit{a} is the lattice parameter) of this local
lattice distortion has an order of $10^{ - 4}$ [2]. There is also a general
ferroelastic lattice distortion associated with the metal-insulator
transition [2] the amplitude of which is much higher, $\delta \cong 10^{ -
2}$.

The energy of an elementary excitation in a semiconductor is equal to the
band gap $E_{g} $. For semiconductors with similar chemical compositions,
there is a relation between the band gap at zero temperature and the energy
$E_{v} $ of the vacancy formation which is given by the equation similar to
the equation (\ref{eq1}),

\begin{equation}
\label{eq5}
E_{v} = \alpha k_{B} T_{m} ,
\end{equation}

\noindent
where $T_{m} $ is the melting temperature. For cadmium chalcogenides, the
ratio $E_{g} /E_{v} $ is $E_{g} /E_{v} \approx 0.8$. For zinc chalcogenides,
this ratio is $E_{g} /E_{v} \approx 1.15$. In GaAs and GaSb, the band gap is
$E_{g} \approx 0.6E_{v} $.

\begin{center}
---------------------------------------------------------------
\end{center}

[1] F.V.Prigara, arXiv:0805.4325 (2008).

[2] F.V.Prigara, arXiv:0811.1131 (2008).

\end{document}